\documentclass[prb,twocolumn,showpacs,preprintnumbers,amsmath,amssymb,superscriptaddress]{revtex4}
\usepackage{graphicx}
\usepackage{dcolumn}
\usepackage{bm}
\usepackage{color}
\usepackage{subfigure}
\usepackage{xfrac}
\usepackage{epstopdf}

\begin{document}

\title{Pressure-Temperature phase diagram of multiferroic EuTiO$_3$}

\author{P. Parisiades}\email{parisiad@esrf.fr}
\affiliation{ID27 Beamline, European Synchrotron Radiation Facility, 71 Avenue des Martyrs, 38000 Grenoble, France}
\author{E. Liarokapis}
\affiliation{Department of Physics, National Technical University of Athens, GR-15780 Athens, Greece}
\author{J. K\"{o}hler}
\affiliation{Max-Planck-Institute for Solid State Research, Heisenbergstr. 1, D-70569 Stuttgart, Germany}
\author{A. Bussmann-Holder}
\affiliation{Max-Planck-Institute for Solid State Research, Heisenbergstr. 1, D-70569 Stuttgart, Germany}
\author{M. Mezouar}
\affiliation{ID27 Beamline, European Synchrotron Radiation Facility, 71 Avenue des Martyrs, 38000 Grenoble, France}
\vskip 0.7cm

%Crystallographic aspects of phase transformations; pressure effects 61.50.Ks
%Transition temperature variations, phase diagrams 74.62.-c
%Transition temperature variations, phase diagrams - Effects of pressure 74.62.Fj

\begin{abstract}

The structural transformation of multiferroic EuTiO$_3$ has been intensively investigated by synchrotron x-ray diffraction at pressures up to 50.3 GPa and temperatures from 50 to 500 K. An antiferrodistortive phase transition from cubic \textit{Pm-3m} to tetragonal \textit{I4/mcm} space group has been observed, identical to the one that has been previously explored at ambient pressure and low temperatures. Several compression/decompression cycles at different temperatures have been carried out to accurately map the transition, and as a result a P-T phase diagram for EuTiO$_3$ has been constructed. The observed phase transition exhibits many similarities with isostructural SrTiO$_3$, although the absence of magnetoelectric interactions in the latter accounts for the different phase boundaries between the two materials.

\end{abstract}

\maketitle

\section{INTRODUCTION}

Multiferroics have been a subject of intense study in recent years, because of the complexity of correlations between lattice, magnetism, strain, and ferroelectric polarization, as well as their potential for practical applications as sensors, spintronics and nonvolatile memories \cite{Nan, Martin, Yeh, Scott}. Europium titanate (EuTiO$_3$) is a multiferroic material at low temperatures with an unusually strong spin-lattice coupling \cite{Shvartsman, Birol}. The strongly localized \textit{4f} moments at  the Eu$^{2+}$ sites order antiferromagnetically at T$_N$ = 5.5 K and are arranged in a G-type structure \cite{McGuire, Katsufuji, Scagnoli}. The dielectric constant decreases abruptly at the onset of antiferromagnetic order \cite{Katsufuji}, and recovers in a magnetic field providing thus direct evidence for strong magnetoelectric coupling at low temperatures.

EuTiO$_3$ adopts the same cubic perovskite structure as SrTiO$_3$, crystallizing in the \textit{Pm-3m} space group. An antiferrodistortive transition to the tetragonal \textit{I4/mcm} structure takes place in both systems on cooling, and is attributed to the TiO$_6$ octahedral rotational instability. For EuTiO$_3$, this tetragonal distortion has first been detected by an anomaly in the specific heat close to room temperature (282 K)\cite{Bussmann} and has been confirmed by other techniques, namely x-ray diffraction\cite{Allieta, Kennedy, Kim, Goian}, EPR\cite{EPR}, muon spin rotation\cite{Guguchia}, resonant ultrasound scattering\cite{Spalek}, inelastic neutron scattering\cite{Ellis}, thermal expansion\cite{Reuvekamp}, magnetic susceptibility\cite{Caslin}. Among various investigations of this phase transition a strong scattering in the transition temperature (T$_s$) has been reported ranging from 160 to 308 K \cite{Bussmann, Allieta, Kennedy, Kim, Goian, Spalek, Ellis}, depending on the applied experimental technique, the fabrication method and the quality of the samples.

The  low temperature antiferromagnetic phase of EuTiO$_3$ has been intensively explored in the last few years \cite{Allieta, Kennedy, Kim, Spalek}, however, the effect of high pressure at high and low temperature is unexplored, with the exception of a pressure dependent study of T$_N$ up to 6 GPa \cite{Guguchia}, where a nonlinear enhancement of the Neel temperature with increasing pressure was reported. On the other hand, high pressure studies of the isostructural SrTiO$_3$ \cite{Guennou, Grzechnik, Salje} have revealed a transition to a tetragonal structure which is identical to the low temperature phase for T$\leq$105K \cite{Muller, Shirane}. The structural similarities between SrTiO$_3$ and EuTiO$_3$ suggest that both materials can exhibit analogous behavior under hydrostatic pressure. In this work we have investigated in detail both the effects of temperature and pressure on the structure of EuTiO$_3$ by synchrotron x-ray diffraction with the purpose to establish its P-T phase diagram and to explore the possibility of any further transitions in an extended pressure-temperature range as compared to previous studies.

\section{EXPERIMENTAL}

Polycrystalline samples of EuTiO$_3$ have been prepared as described in Ref\cite{Bussmann}. The samples were checked by x-ray diffraction and specific heat and showed cubic symmetry at room temperature and a specific heat anomaly at T=282 K as expected for the reported phase transition.

The synchrotron x-ray diffraction experiments were performed on the ID27 beamline at the ESRF. Monochromatic wavelengths of 0.24678 and 0.3738 \AA were selected. The beam size was focused to a spot of about 3x3 $\mu$m$^2$. The samples were loaded in Le Toullec type membrane diamond anvil cells (DACs) with diamond culets of 300 $\mu$m in diameter. The rhenium gaskets were pre-indented to a thickness of 50 $\mu$m, while the sample chamber was created by drilling a 150 $\mu$m hole in the gasket with a Nd:YAG pulsed laser.

The proximity of the cubic-to-tetragonal transition of EuTiO$_3$ at ambient conditions requires systematic measurements at both low and high temperatures in order to sufficiently explore the high pressure behavior of this system. Thus, several pressure ramps at different temperatures (50, 200, 295, 400 and 500 K) were carried out up to a maximum pressure of 50.3 (1) GPa. For room and high temperature measurements, Ne has been used as a pressure transmitting medium. The high temperature data were obtained with an external resistive heating device and the temperature was measured by a thermocouple that was in contact with the back side of the diamond. For the low temperature datasets, a liquid helium flow cryostat has been used and the cells were gas-loaded with He. Both pressure media provide very good hydrostatic conditions within the pressure range of interest \cite{Klotz}. The pressure has been measured using the fluorescence lines of ruby at room and low temperatures and those of SrB$_4$O$_7$:Sm$^{2+}$ for higher temperatures. The data were collected with Mar image plate and Perkin Elmer flat panel detector and the acquired images have been integrated using the Fit2D software \cite{Hammersley}. The diffractograms were indexed using DICVOL91 software \cite{Boultif}. The Lebail refinements have been carried out with the Fullprof software package \cite{Carvajal}.

\section{RESULTS}

At ambient conditions, the EuTiO$_3$ x-ray diffraction pattern can be well refined by the cubic space group \textit{Pm-3m}, with a = 3.9046 (1) \AA, in good agreement with previous works \cite{Allieta, Kennedy, Kohler, Ellis2}. On cooling, EuTiO$_3$ undergoes a cubic-to-tetragonal structural transition to the \textit{I4/mcm} space group. The transition temperature varies among the different studies: specific heat, thermal expansion, inelastic neutron scattering measurements identify the transition temperature as T$_S$=282$\pm$3 K \cite{Bussmann, Reuvekamp, Ellis}, and a coupling of the R-point order parameter with macroscopic strains has been proposed by resonant ultrasound spectroscopy at 284 K \cite{Spalek}. On the other hand, various diffraction experiments find a lower transition temperature in the range between 160-240 K \cite{Allieta, Kennedy, Kim} whereas Goian et al\cite{Goian} report the phase transformation near 300 K for samples sintered at ambient pressure. For samples sintered at high pressure no transition was observed\cite{Goian}. This suggests that the inherent microstrain induced during the fabrication process may play a large role in the determination of the transition boundaries. Another scenario that has been proposed through careful pair distribution function analysis of the x-ray data, suggesting the presence of local tetragonal regions inside a long-range cubic phase, which evolve to the bulk as the temperature is further decreased\cite{Allieta}.

\begin{figure*}
\centering
\subfigure{\includegraphics[width=0.45\linewidth]{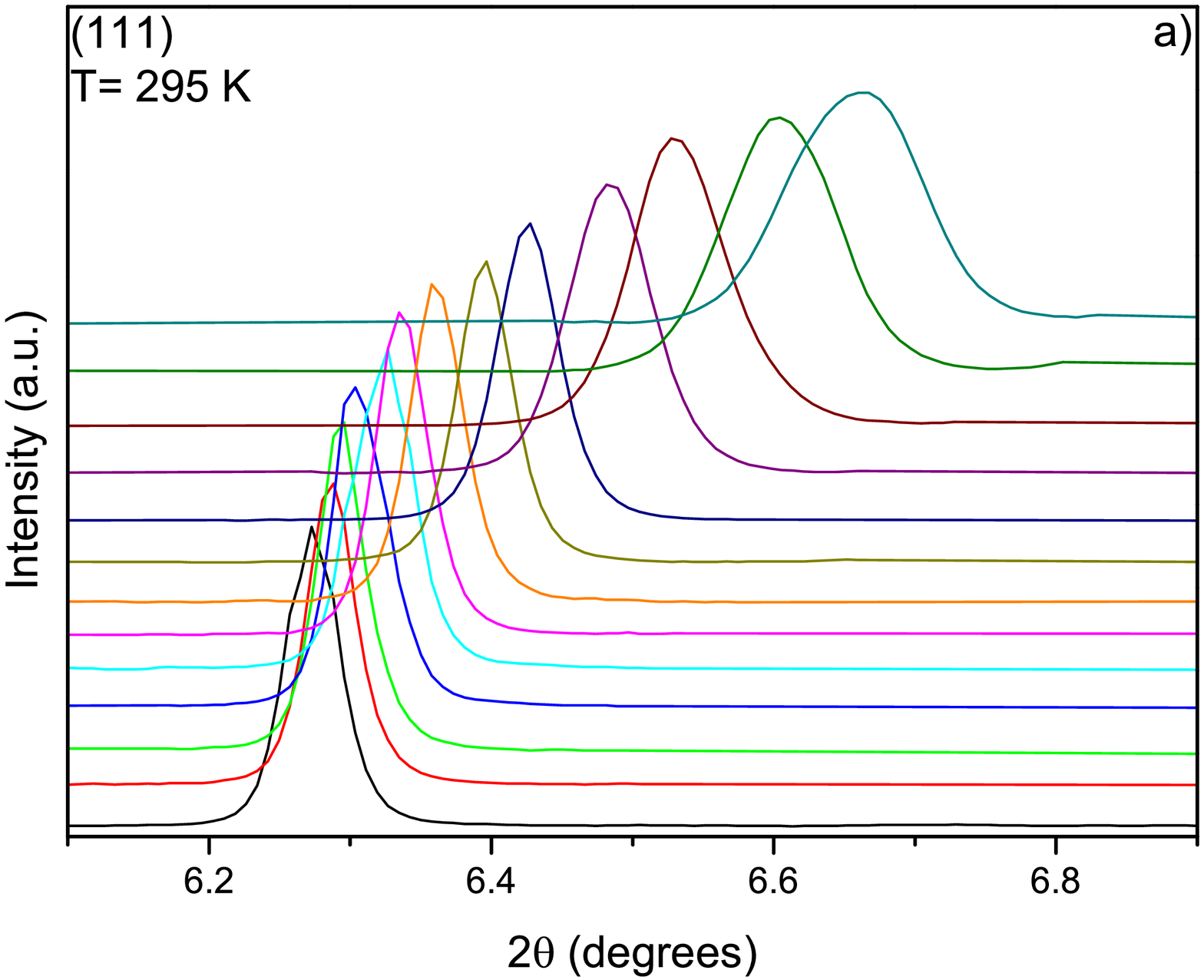}}
\subfigure{\includegraphics[width=0.4725\linewidth]{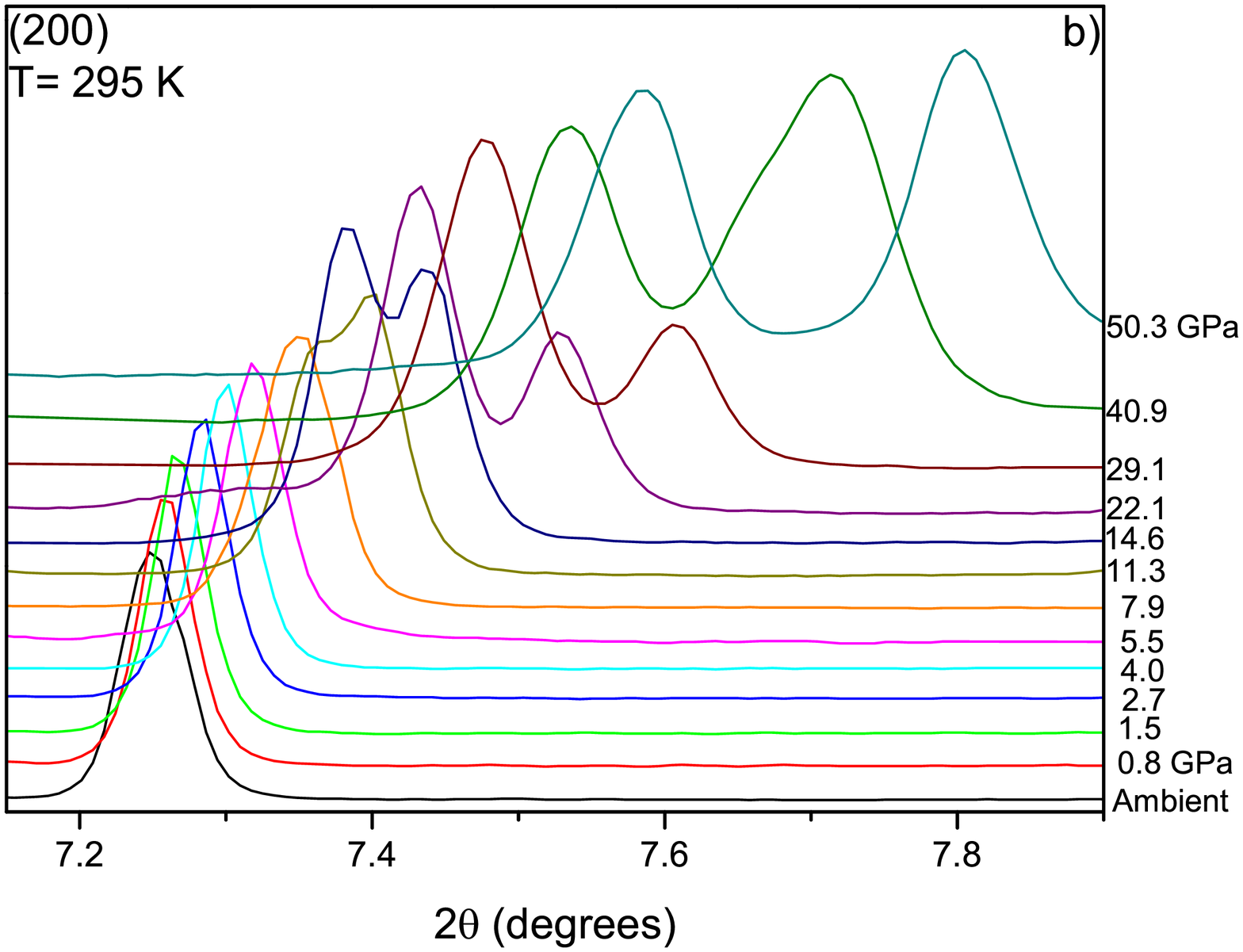}}
\caption{(Color online) Fig.1 Selected high pressure data at room temperature for EuTiO$_3$. Evolution of the a) (111) and b) (200) reflection with pressure. The splitting in (200) is indicative of the transition from the cubic \textit{Pm-3m} to the tetragonal \textit{I4/mcm} structure. The data were collected using a wavelength of 0.24678 \AA. }
\end{figure*}

In Fig.1 we present some selected high pressure synchrotron diffraction data at room temperature. A large number of reflections, such as the (200), split into a doublet with increasing pressure while the (111) peak remains unperturbed up to the highest pressure of 50.3 (1) GPa studied here. This behavior is fully consistent with the one observed in the low temperature datasets \cite{Allieta, Kennedy}, and the data can indeed be fitted by the tetragonal \textit{I4/mcm} space group for higher pressures. Other possible structures such as the \textit{Imma}, \textit{R-3c}, and \textit{I4/mmm} space groups were excluded since the refinement model is not in agreement with any of the acquired data. Some additional line broadening is also expected because of the Ne PTM above 20 GPa \cite{Klotz}. The transition is fully reversible, as shown by the decompression data (presented in Supplemental Material (SM), Part 1), although the number of points upon releasing pressure are not sufficient to detect any possible hysteresis. In Figs.2a,b we present the Lebail refinements at 1.5 (1) and 12.9 (1) GPa, fitted with the cubic and tetragonal model, respectively. The observation of the phase transition is also confirmed by the comparison of the (200) peak which splits into (004) and (220) in the tetragonal phase (see inset). This cubic-to-tetragonal transition has been shown to be driven by the out-of-phase rotations of the corner-sharing octahedra \cite{Bettis} and can also be generated by a Jahn-Teller distortion, which is common for ABO$_3$ perovskites \cite{Carpenter}. This rotational instability of the TiO$_6$ octahedra is also present in the isostructural SrTiO$_3$, and is responsible for the temperature- and pressure-induced transitions to the tetragonal \textit{I4/mcm} phase for this material \cite{Guennou, Salje}. The tilting of octahedra is accompanied by the softening of a transverse acoustic zone-boundary mode, and has the same temperature dependence in both SrTiO$_3$ and EuTiO$_3$ \cite{Bettis}. During the preparation of this paper, we became aware of another high pressure study at room temperature that has been carried out up to pressures of 30 GPa \cite{Syassen}. In this study the lattice parameters have been refined as a function of pressure and the critical pressure for the structural phase transition was also defined.

\begin{figure*}
\centering
\subfigure{\includegraphics[width=0.45\linewidth]{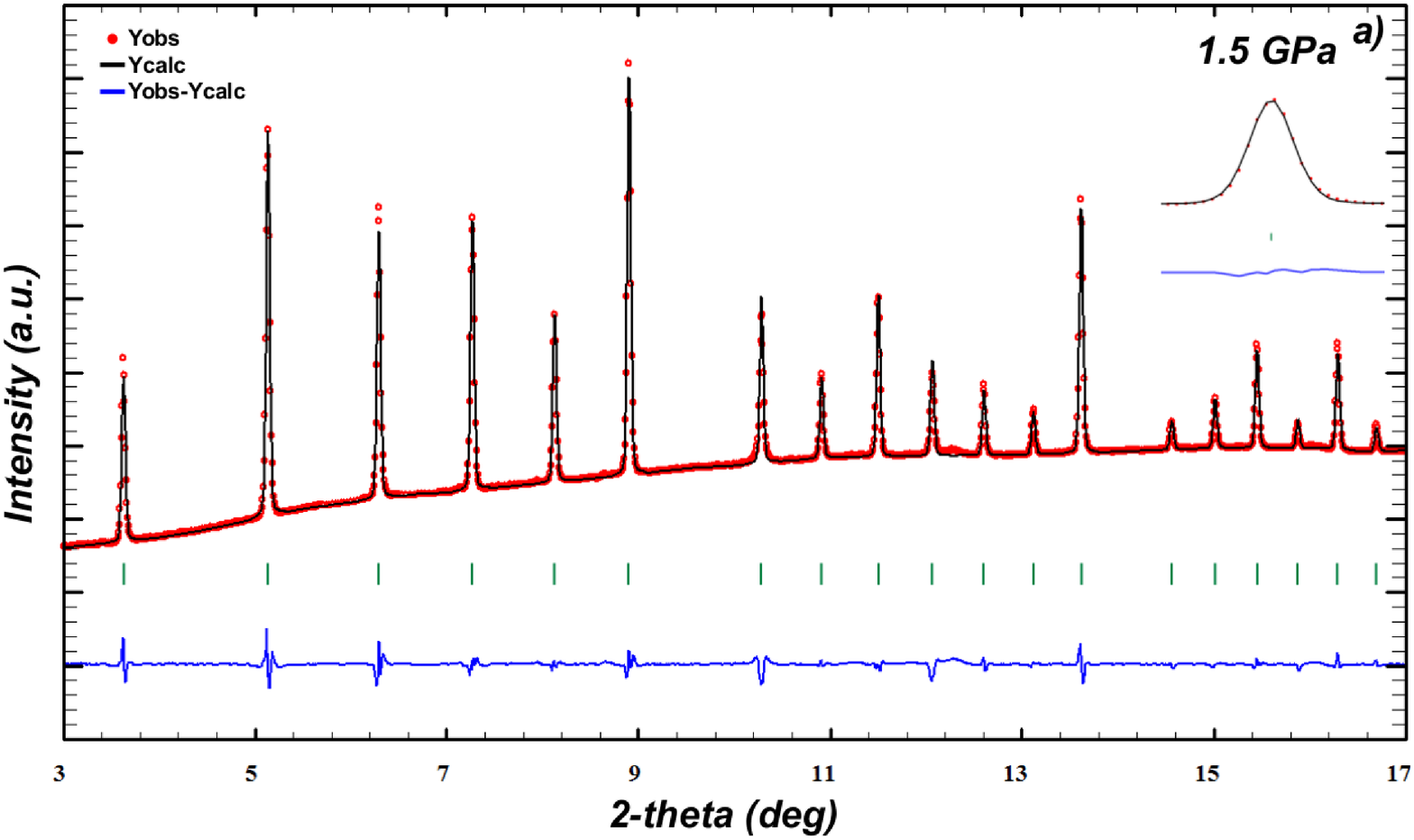}}
\subfigure{\includegraphics[width=0.45\linewidth]{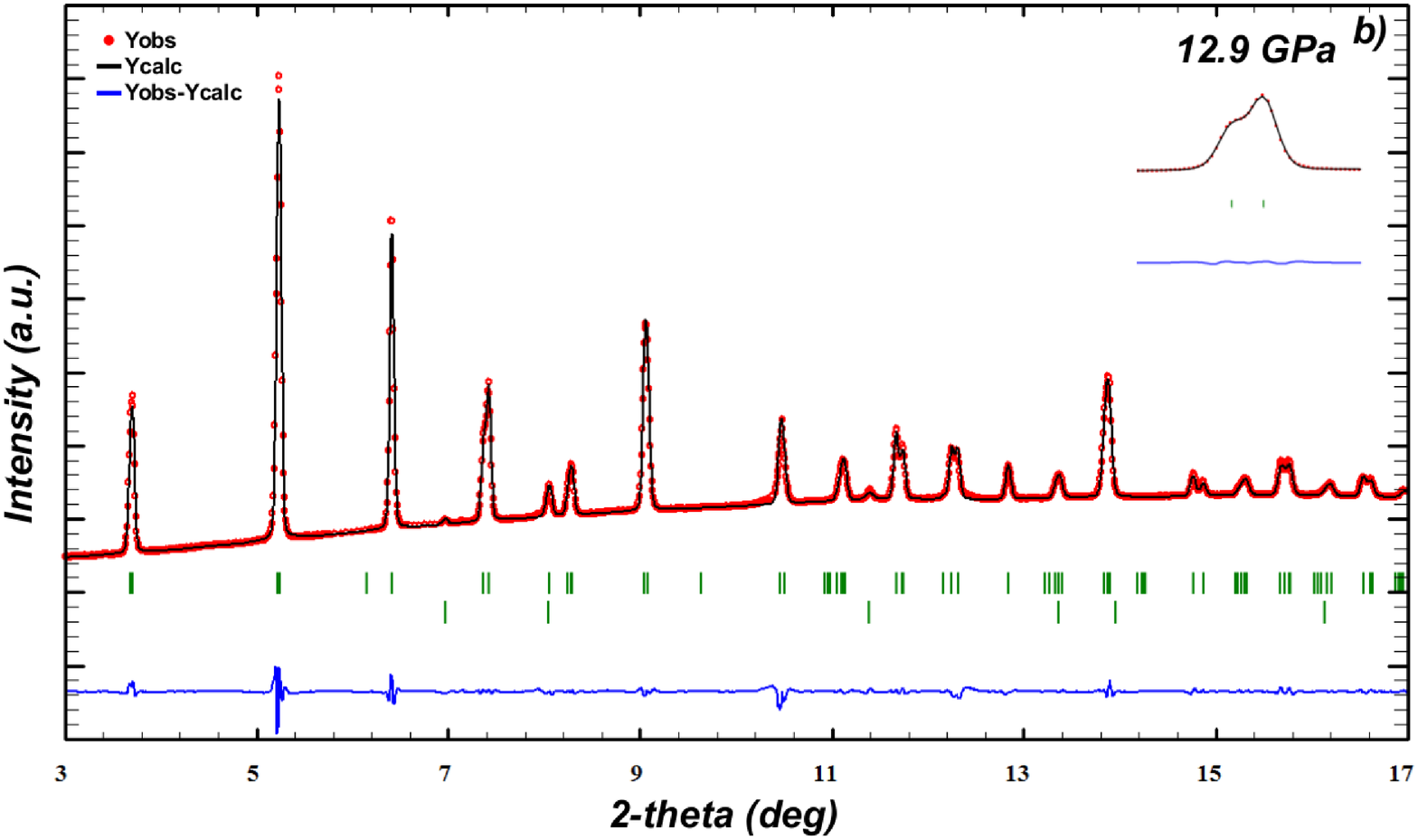}}
\caption{(Color online) Lebail refinements of EuTiO$_3$ for a) the cubic \textit{Pm-3m} phase at 1.5 (1) GPa and for b) the tetragonal \textit{I4/mcm} phase at 12.9 (1) GPa. Inset in a) shows the single (200) reflection in the cubic phase and in b) shows the splitting of the (200) reflection to (220) and (004) in the tetragonal phase. The secondary phase in Fig.2b corresponds to peaks arising from the solidification of Ne transmitting medium.}
\end{figure*}

In order to accurately define the boundaries of the observed transition, we performed several additional high pressure experiments, at both low and high temperatures. For the low temperature measurements, a DAC loaded at an initial pressure of $\sim$1.1 GPa has been cooled down to 40 K. A splitting of the (h00) family of reflections occurs on cooling, indicative of the antiferrodistortive transition \cite{Allieta, Kennedy}. As expected, the transition temperature is shifted with pressure. The lattice changes on cooling are much more subtle than on compression, since the distortions from the cubic unit cell are small, but splitting of the (400) Bragg peak is clearly visible at 180 K (See SM, Part 2). Upon increasing the pressure at 200 K and 50 K up to 18.8 and 19.6 GPa, respectively, the tetragonal structure is well resolved. The increase of pressure at 200 K further enhances the visibility of the transition and at higher pressures the majority of the peaks has split, analogous to the room temperature experiment (See SM, Part 3). A careful examination of the data did not reveal any additional transitions in this low-temperature and high-pressure region. The same is true for the compression at 50 K, with the compound remaining in the tetragonal \textit{I4/mcm} phase for the whole pressure range (See SM, Part 4).

High temperature compressions at 400 K and 500 K were also performed up to 38.6 and 31.1 GPa (some selected diffraction data are shown in SM, Parts 5\&6). For all experiments, the onset of the transition has been marked by the pressure dependence of the normalized full-width at half-maximum (FWHM) for the (200) cubic reflection for 295, 400 and 500 K, as shown in Fig.3. Deviatoric stresses can be generated during compression at RT and LT but these stresses are released upon heating (as indicated by the sharper peaks at high temperatures). By using a single profile function for the simulation with a cubic phase, the FWHM of the (200) reflection increases at different critical pressures for each temperature. It has a linear behavior at high temperatures, while it exhibits a more abrupt increase at room temperature. At ambient temperature the sample adopts the tetragonal space group with a = 5.49304 \AA\ (2) and c = 7.77206 (2) \AA\ and the onset of the transition is at 2.7 GPa according to the (200) FWHM increase. The transition pressures are 5.7 and 10.5 GPa at 400 and 500 K respectively. In the inset of Fig.3, the FWHM evolution of the (200) Bragg peak at low temperatures is shown, with a broadening that occurs at 240 K for 1.1 GPa.

\begin{figure}
\centering
\includegraphics[width=0.8\linewidth]{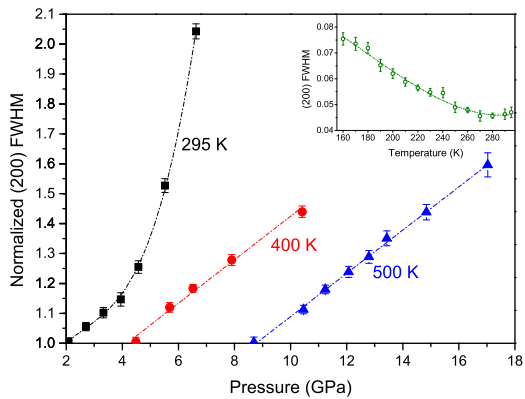}
\caption{(Color online) Evolution of the FWHM of the (200) Bragg reflection with pressure at temperatures 295 (black rectangles), 400 (red circles) and 500 K (blue triangles). The critical pressures for the transition at the three temperatures are defined as 2.7, 5.5 and 10.5 GPa, respectively. In the inset the FWHM of the (200) peak at low temperatures and 1.1 GPa is presented, with the transition temperature being at 240 K.}
\end{figure}

The pressure-volume relationship for EuTiO$_3$ at room temperature is displayed in Fig.4a. The fits for the two observed phases have been carried out using the Murnaghan (M) equation of state (EoS) for the low pressure cubic phase, while the 3rd order Birch-Munaghan (BM3) EoS has been employed for the high pressure tetragonal structure. The modeling yields bulk moduli of 180.9 and 190.3 (8) GPa for the two phases at room temperature. Since there are no previous high pressure data, we can only compare the results with some recent DFT calculations for the two phases, which provide 162.5 and 171.1 GPa \cite{Rushchanskii} for the cubic and the tetragonal lattice, respectively. Our results are in good agreement with theory, especially by considering the uncertainty of DFT in the calculation of lattice constants. Experiments on SrTiO$_3$ yield a bulk modulus between 165-176 GPa for the cubic phase \cite{Guennou, Ishidate, Edwards, Beattie}, comparable to EuTiO$_3$ while in the tetragonal phase it is somewhat larger (225 GPa) \cite{Guennou}. The determination of the critical pressures for the transition can play a significant role in determining the bulk moduli. For example, different onsets are established by defining the transition by a FWHM increase/peak splitting as compared to the appearance of superstructure reflections. The pressure dependence of the structural parameters is also shown in Fig.4b, where we have used the converted lattice constants for the tetragonal structure (\textit{a=a$_T$/$\sqrt{2}$}, \textit{c=c$_T$}/2 ). The lattice parameters decrease monotonically in the entire range of measurements, with the a-axis decrease being stronger than the c-axis related one. This behaviour is in accordance with the data of Ref.\cite{Syassen} at room temperature, however with the lattice parameters being slightly smaller than here. The tetragonal distortion is very weak at low pressures, as is also evident from the \textit{c/a} ratio (SM, Part 7), but the FWHM evolution provides a reasonable measure about the onset of the transition, which is lower than the one deduced for the same sample from specific heat measurements \cite{Bussmann}. A similar situation is observed for SrTiO$_3$, where the a-axis is more sensitive to pressure in the tetragonal phase \cite{Guennou}.

\begin{figure}
\centering
\includegraphics[width=0.8\linewidth]{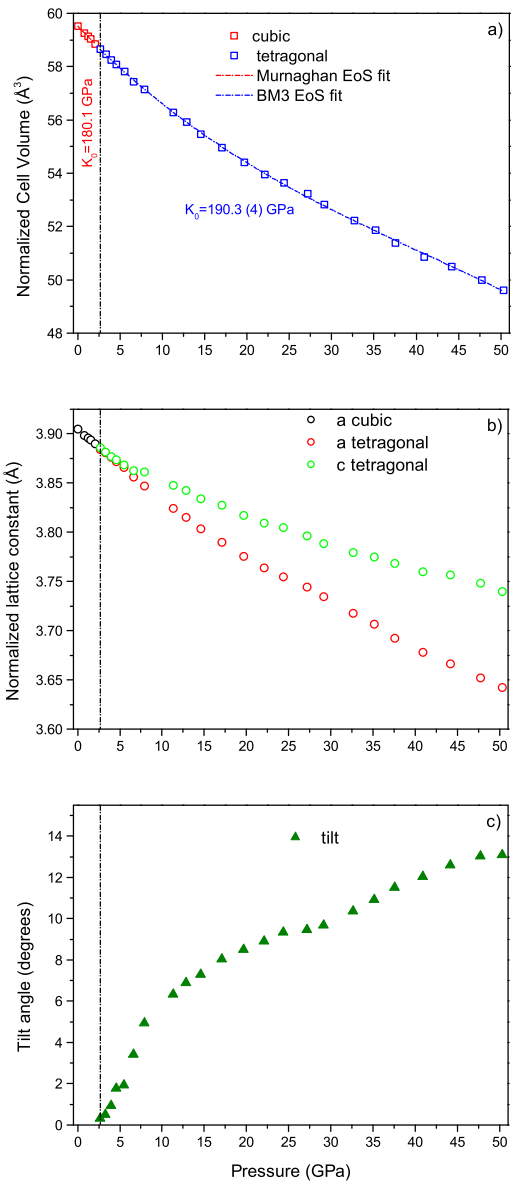}
\caption{(Color online) a) V-P relationship, b) normalized lattice parameters and c) octahedral tilt angle for EuTiO$_3$ at ambient temperature. In all figures, the error bars are smaller than the symbol size.}
\end{figure}

In the tetragonal \textit{I4/mcm} phase (tilt system a$^0$a$^0$c$^-$ in Glazer notation\cite{Glazer}, there is one out-of-phase rotation along the c-axis and no rotations along the a- and b- axes. This results in displacements of the O atoms (O2 site) from (\sfrac{1}{4}, \sfrac{3}{4}, 0) to (\sfrac{1}{4}+u, \sfrac{3}{4}+u, 0). The polycrystallinity of our samples did not allow for a complete Rietveld refinement in order to determine the exact atomic positions. However, for the case of rigid TiO$_6$ octahedra (maintaining the corner-sharing connectivity), an alternative expression of the tilt angle $\phi$ in the tetragonal phase can be calculated via the relation $\phi=arccos(\sqrt{2}a/c)$, a and c being the lattice parameters. This is shown in Fig.4c. In the case of EuTiO$_3$, the tilt angle corresponds to the deviation from the cubic structure and represents the main order parameter of the transition.

\section{DISCUSSION}

By collecting all the data for the antiferrodistortive phase transition we present first P-T phase diagram for EuTiO$_3$ in Fig.5. Some structural data from other work at ambient pressure are included for comparison. To a good approximation our data can be fitted by a straight line, and the extrapolation of the linear fit yields a transition temperature of 221 K at ambient pressure, with a slope of \textit{dP$_c$/dT}=0.036 GPa/K, \textit{P$_c$} being the critical pressure of the transition. As mentioned above, the transition temperature at ambient conditions can vary from one experiment to the other. Oxygen deficiencies can be an important factor for these small discrepancies, a fact that can also be attributed to the small variation of the lattice parameters at ambient conditions among the different studies. An alternative defect can also be the presence of Eu$^{3+}$ ions, which have smaller ionic radius than Eu$^{2+}$, resulting in  reduced lattice parameters, which also affect the magnetic properties. Goian et al \cite{Goian} proposed that the transition temperature T$_s$ in EuTiO$_3$ is strongly dependent on the sample preparation: they did not observe any tetragonal distortion down to 100 K from single crystal diffraction, while powder diffraction of the same sample revealed the tetragonal structure close to room temperature, as observed by an anomaly in the thermal expansion. Eu or Ti vacancies are also possible defects for this system, contrary to SrTiO$_3$ \cite{McCarthy}.

\begin{figure}
\centering
\includegraphics[width=0.8\linewidth]{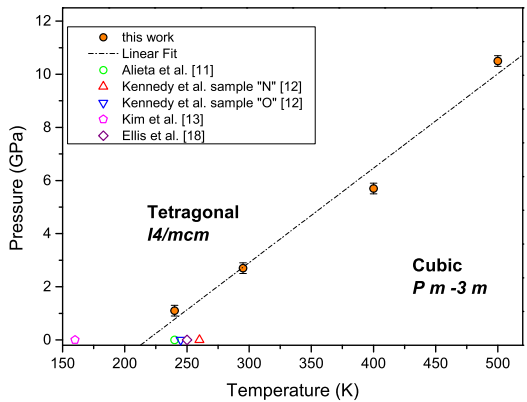}
\caption{(Color online) P-T phase diagram for EuTiO$_3$, together with previously published results (using X-Ray diffraction) for the onset of the tetragonal \textit{I4/mcm} phase at ambient pressure. }
\end{figure}

The x-ray diffraction results for the transition temperature are consistently lower than the results obtained by other techniques, where the transition has been observed much closer to room temperature, such as specific heat \cite{Bussmann} (282 K), inelastic x-ray scattering \cite{Ellis} (287 K) or resonant ultrasound spectroscopy \cite{Spalek} (284 K), thermal expansion \cite{Reuvekamp}, magnetic susceptibility \cite{Caslin}. This difference can be attributed to the correlation length scale evolution of the structural transition, indicating that the transition occurs initially at the nanoscale before being bulk sensitive, as shown by pair distribution function analysis \cite{Allieta}. As long as there is no long range coherence in the structure, it is very difficult to detect minor tetragonal distortions in the structure with XRD, but it is possible to detect them using specific heat, and this can be a reason for the different transition temperatures between the two techniques. To minimize this discrepancy we have defined the onset of the transition by the abrupt (200) peak broadening and not by the splitting of the peaks, since such local effects, if present, should be reflected in the FWHM. However, the reduced resolution inside a DAC, even with the highly hydrostatic pressure media used in this work (He and Ne) cannot be ruled out completely as a source of error and can affect the precise determination of the transition boundaries. Kim et al\cite{Kim} observed an incommensurate and long-range modulated phase transition (denoted as m-AFD in the paper) at 285 K, while the simple AFD transition takes place at 160 K. This modulated AFD order is proposed to appear by the competition between the octahedral rotation and electric polarization and seems to exhibit a temporal hysteresis. Their interpretation has, however, been discarded in several recent papers and been attributed to sample inhomogeneity which also explains the strongly reduced antiferromagnetic transition temperature as compared to other work \cite{Ellis2}.

EuTiO$_3$ has many common properties with isostructural SrTiO$_3$: identical lattice constants and valence for Sr and Eu with similar ionic radii within the ABO$_3$ perovskite structure, as well as the recently found optic mode softening \cite{Goian, Kamba}. It is therefore not surprising that the two materials have similar phase diagrams. At room temperature and under high pressure, SrTiO$_3$ transforms to tetragonal at a critical pressure P$_c$=9.6 GPa. The transition evolves linearly at higher temperatures, with a P$_c$= 15 and 18.7 GPa at 381 and 467 K, respectively \cite{Guennou}. According to the phase diagram, the zero pressure transition takes place at 105 K \cite{Guennou, Muller}. The absence of antiferromagnetic order at low temperatures might be the reason why SrTiO$_3$ transforms to tetragonal at much lower temperatures and higher pressures. Moreover, according to DFT+U calculations \cite{Birol, Bettis}, the octahedral rotations are also affected by the hybridization of Eu \textit{4f} states with Ti \textit{3d} states, which is strong in EuTiO$_3$ and absent in SrTiO$_3$. The coupling between Eu \textit{4f} states and Ti \textit{3d} electrons is very sensitive to local breaking of symmetry and therefore the octahedral tilts will change the covalent bonding strength between Eu and O ions\cite{Whangbo}, making the tetragonal lattice the most favourable energy state.

\section{CONCLUSION}

In summary, we report clear evidence for a pressure- and temperature-induced cubic-to-tetragonal phase transition on EuTiO$_3$. The observed transition results from the out-of-phase tilting of TiO$_6$ octahedra. This is, to our knowledge, the first high pressure and high/low temperature diffraction experiment carried out on this compound. By combining several compression datasets at different temperatures, a phase diagram has been constructed in a wide P-T range. We are convinced that this work will motivate theoretical and experimental groups to investigate further possible correlations between structure, magnetism and ferroelectric properties for this material.

\begin{widetext}
{\bf\Large \center Supplementary Online Material for ``Pressure-Temperature phase diagram of multiferroic EuTiO$_3$''}
\setcounter{figure}{0}
\renewcommand{\thefigure}{S\arabic{figure}}

\begin{figure}
\centering
\includegraphics[width=0.7\linewidth]{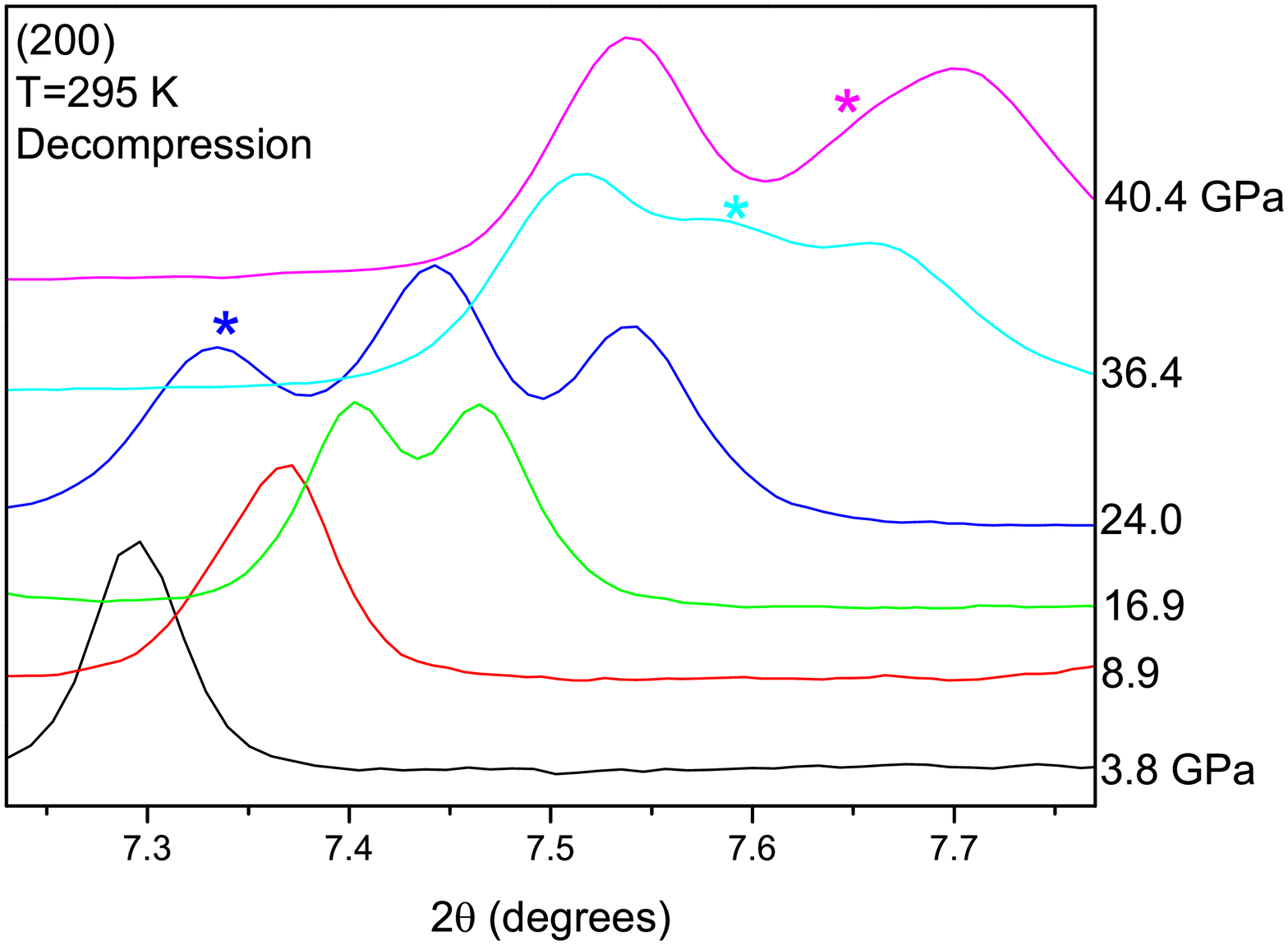}
\caption{(Color online) Evolution of the (200) diffraction peak on decompression at T=295 K. Asterisk marks the diffraction of Ne pressure transmitting medium. The material reverts back to the cubic phase on decompression. The data were collected using a wavelength of 0.24678 \AA.}
\end{figure}

\begin{figure}
\centering
\includegraphics[width=0.7\linewidth]{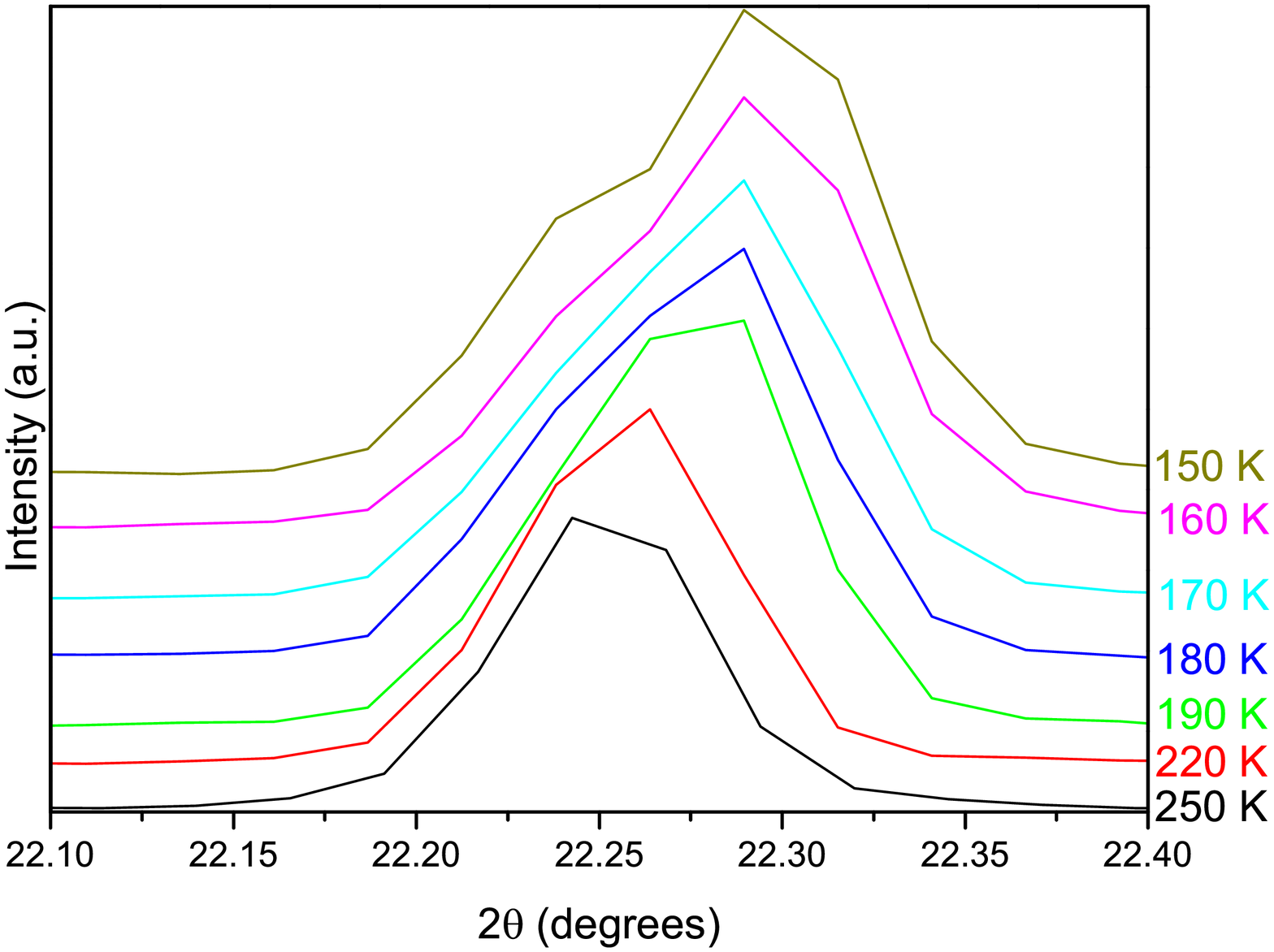}
\caption{(Color online) Evolution of the (400) diffraction peak on cooling at P=1.1 GPa. A shoulder appears at 180 K, with a more visible split at 160 K. The data were collected using a wavelength of 0.3738 \AA.}
\end{figure}

\begin{figure}
\centering
\includegraphics[width=0.7\linewidth]{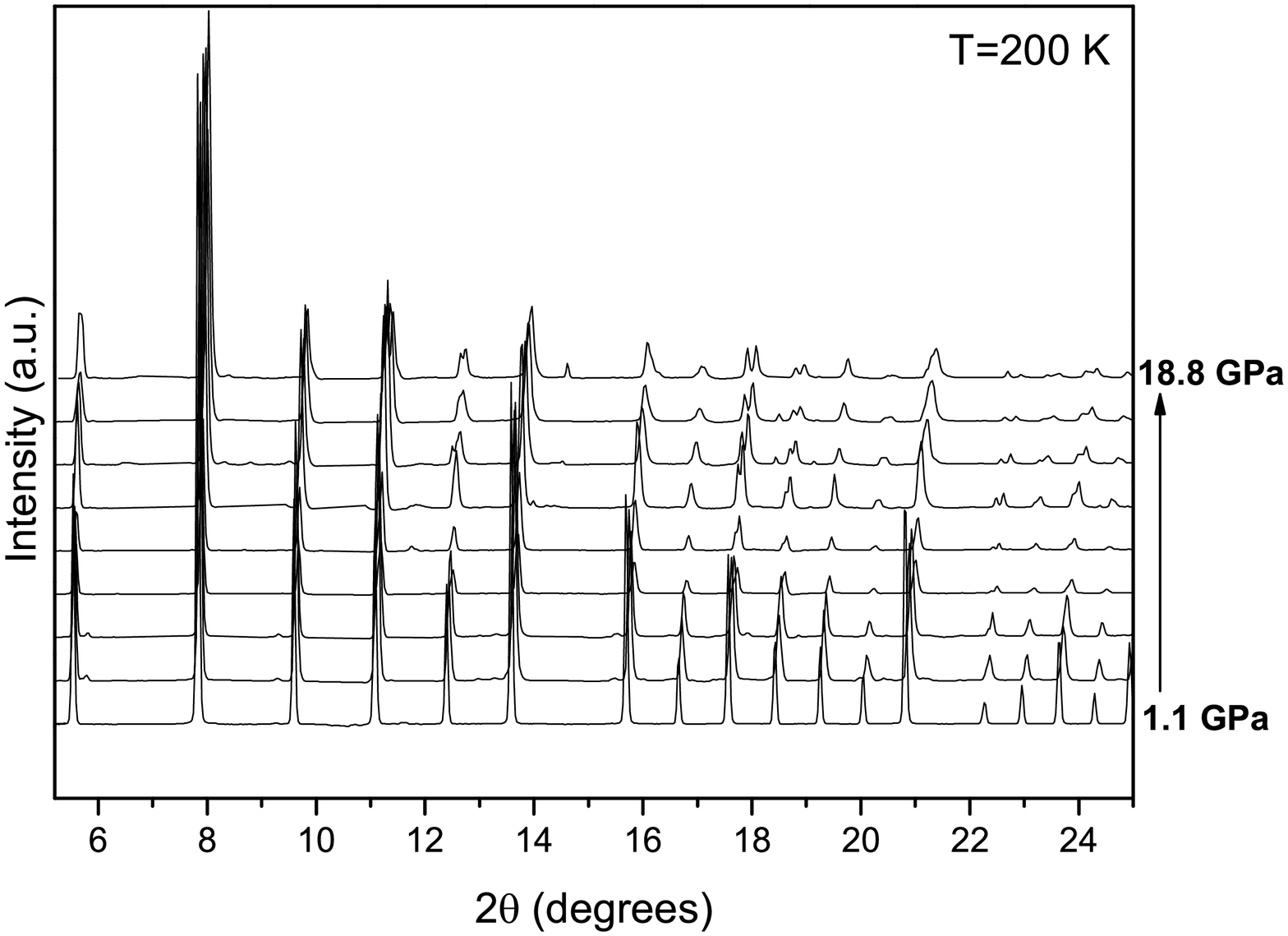}
\caption{Selected compression diffraction data collected at 200 K. The data were collected using a wavelength of 0.3738 \AA.}
\end{figure}

\begin{figure}
\centering
\includegraphics[width=0.7\linewidth]{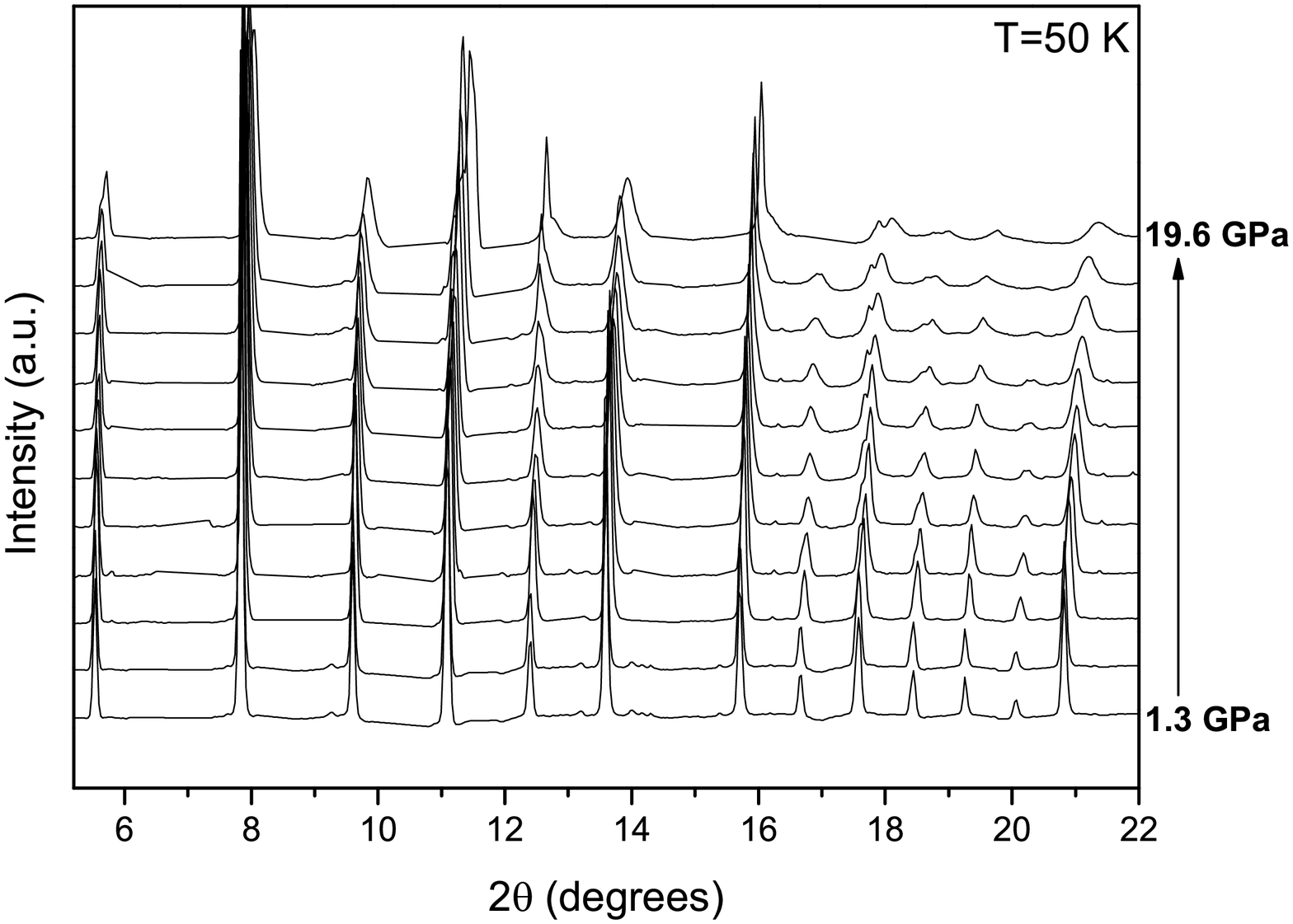}
\caption{Selected compression diffraction data collected at 50 K. The resolution is reduced due to the previously acquired compression-decompression cycle at 200 K, thus the peaks appear broader. The data were collected using a wavelength of 0.3738 \AA.}
\end{figure}

\begin{figure}
\centering
\includegraphics[width=0.7\linewidth]{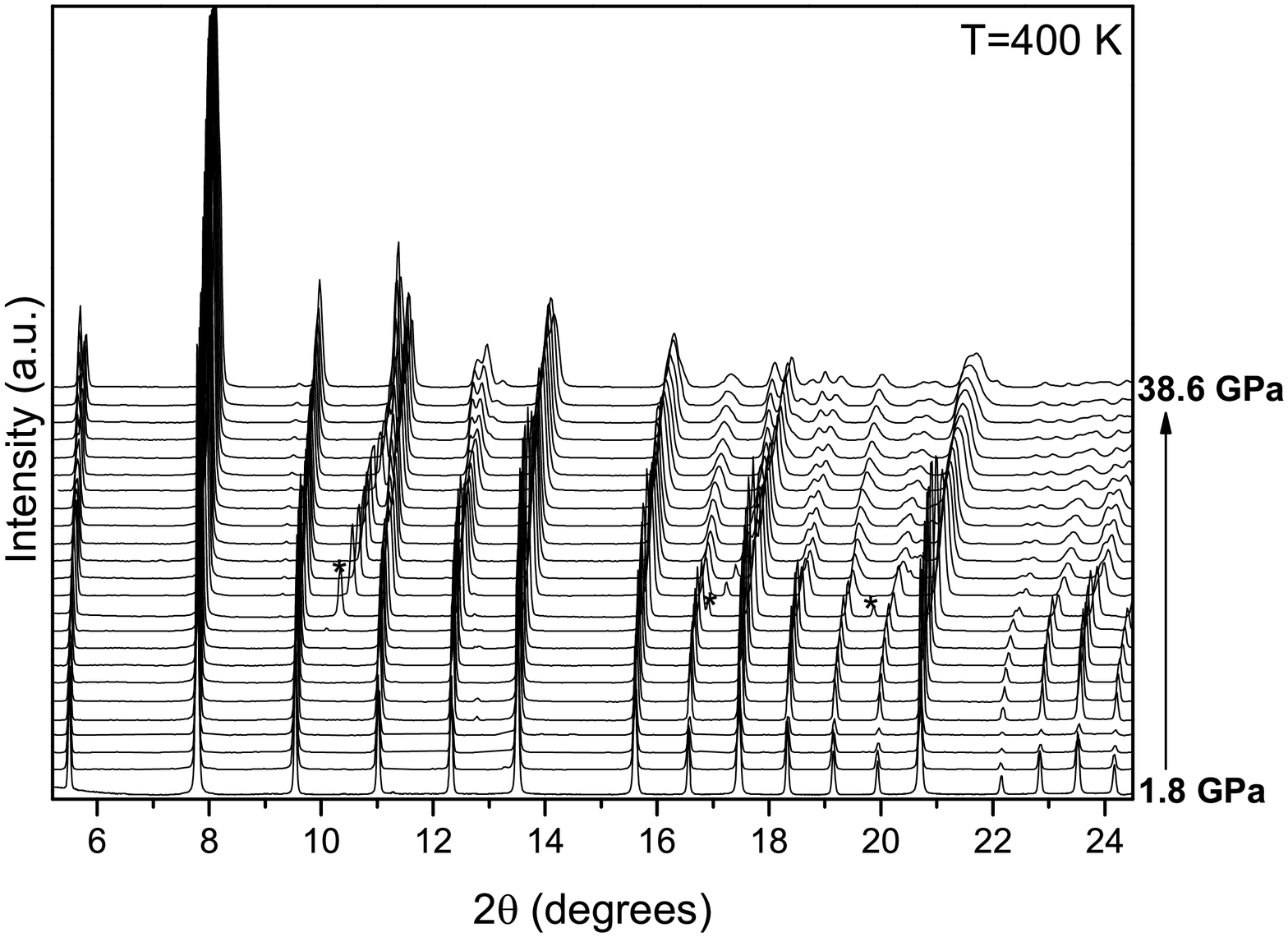}
\caption{Selected compression diffraction data collected at 400 K. Asterisks denote the peaks arising from Neon PTM solidification. The data were collected using a wavelength of 0.3738 \AA.}
\end{figure}

\begin{figure}
\centering
\includegraphics[width=0.7\linewidth]{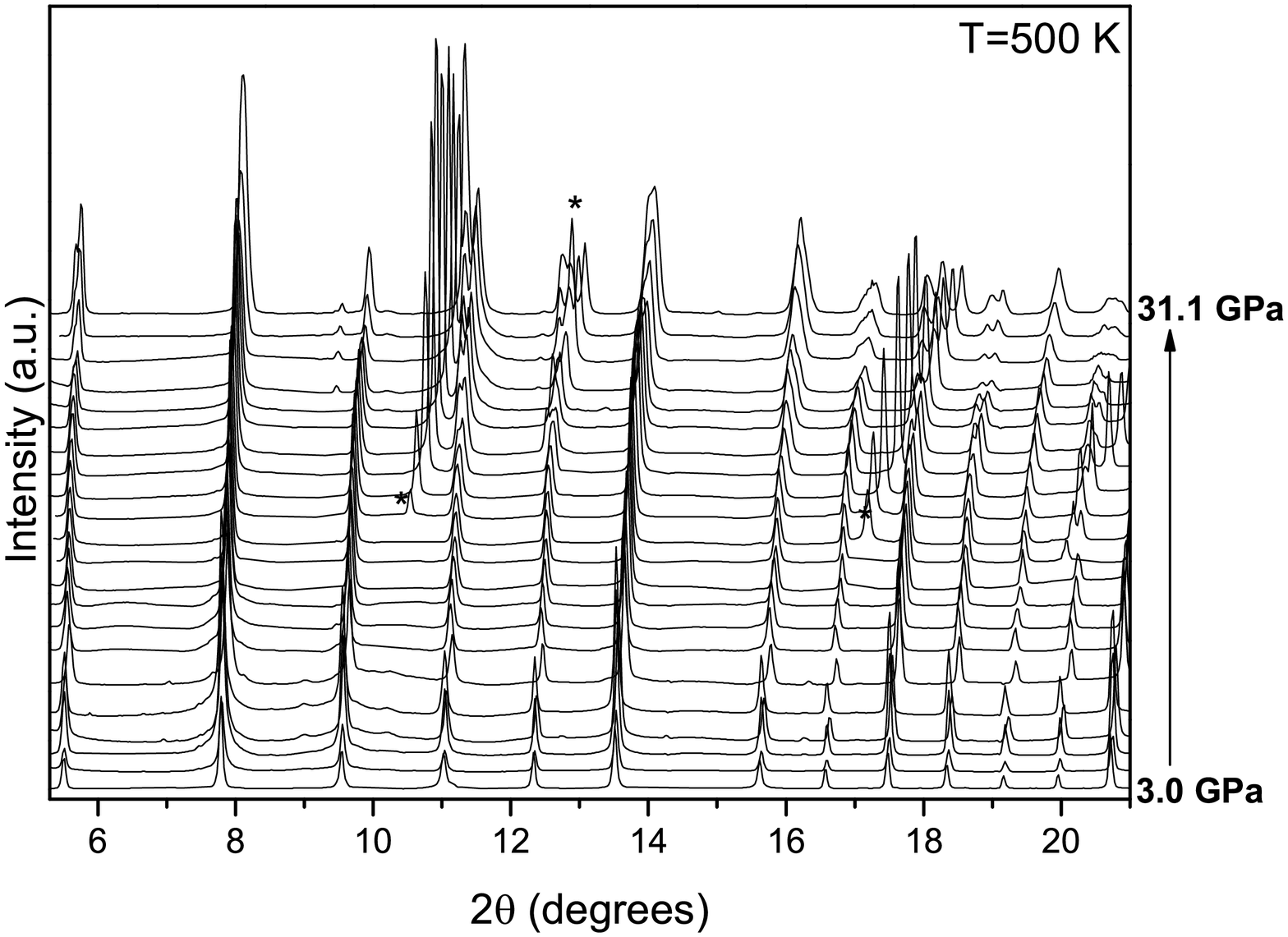}
\caption{Selected compression diffraction data collected at 500 K. Asterisks denote the peaks arising from Neon PTM solidification. The data were collected using a wavelength of 0.3738 \AA.}
\end{figure}

\begin{figure}
\centering
\includegraphics[width=0.7\linewidth]{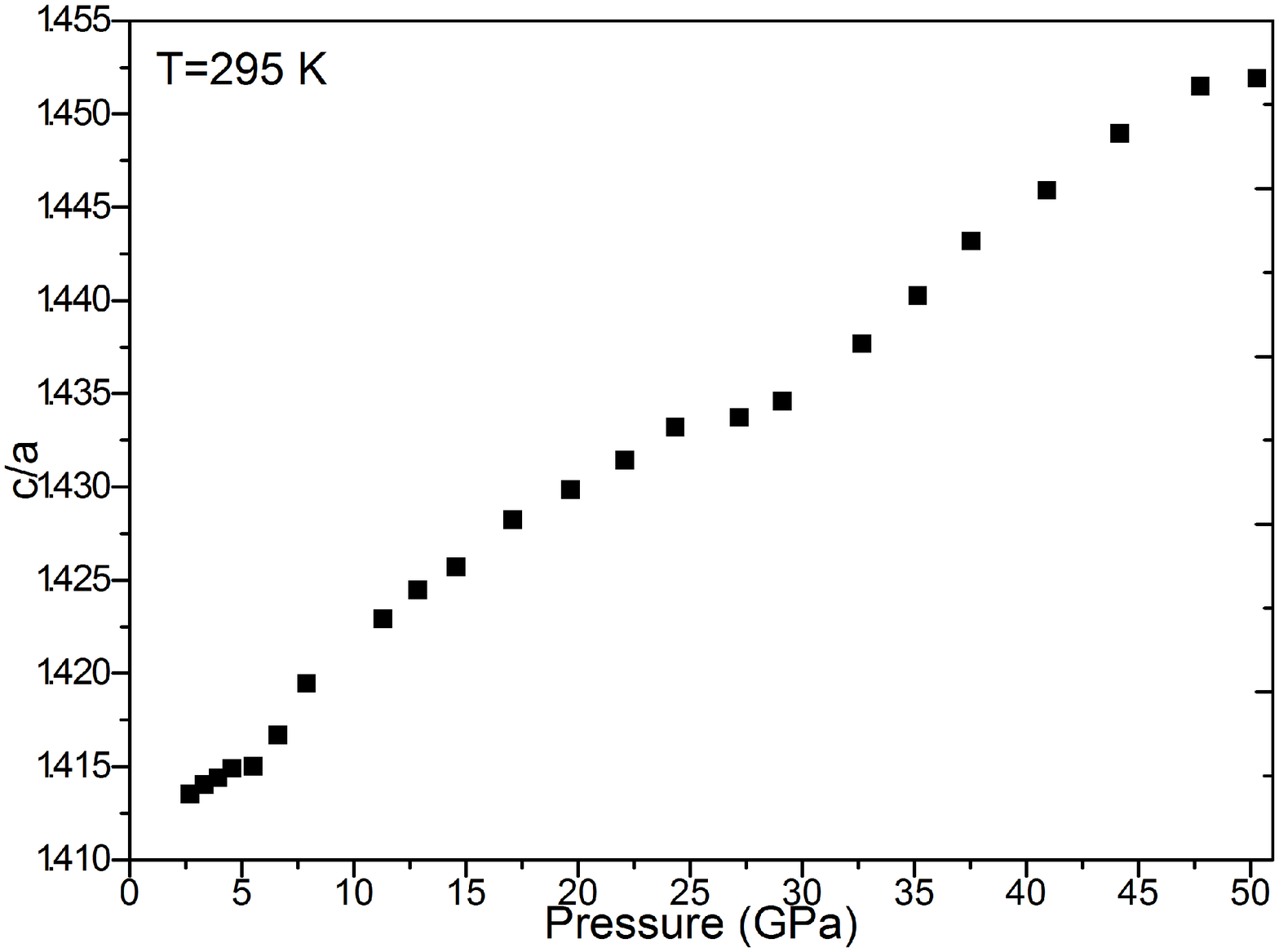}
\caption{c/a ratio for the tetragonal phase of EuTiO$_3$.}
\end{figure}

\end{widetext}

\end{document}